\input harvmac
\input epsf

\def\frak#1#2{{\textstyle{{#1}\over{#2}}}}

\def\ga{\gamma}

\def\Btilde{\tilde B}

\def\Etilde{\tilde E}
\def\Ttilde{\tilde T}

\def\sic{supersymmetric}
   
\def\ssm{supersymmetric standard model}

\def\smgroup{$SU_3\otimes\ SU_2\otimes\ U_1$}

\def\npb{{Nucl.\ Phys.\ }{\bf B}}
\def\plb{{Phys.\ Lett.\ }{\bf B}}
\def\prd{{Phys.\ Rev.\ }{\bf D}}
\def\prl{Phys.\ Rev.\ Lett.\ }
\def\zpc{Z.\ Phys.\ {\bf C}}

\def \in{\leftskip = 40 pt\rightskip = 40pt}

\def \out{\leftskip = 0 pt\rightskip = 0pt}

\def\lf{16\pi^2}
\def\llf{(16\pi^2)^2}
\def\lllf{(16\pi^2)^3}

{\nopagenumbers
\line{\hfil LTH 372}
\line{\hfil hep-ph/9605440}
\line{\hfil Revised Version}
\vskip .5in
\centerline{\titlefont The three-loop SSM $\beta$-functions}
\vskip 1in
\centerline{\bf P.M.~Ferreira, I.~Jack and D.R.T.~Jones}
\bigskip
\centerline{\it Department of Mathematical Sciences, 
University of Liverpool, Liverpool L69 3BX, U.K.}
\vskip .3in
We present the supersymmetric standard model three-loop $\beta$-functions 
for gauge and Yukawa couplings and consider the effect of three-loop 
corrections on the standard running coupling analyses. 

\Date{May 1996}}

The unification (or near--unification) 
of the gauge couplings at $M_{G}\approx 10^{16}$~GeV has
catalysed intensive study of  the \ssm\ (SSM).  The evolution of the
Yukawa couplings from $M_Z$ to $M_G$ is also  of interest. For small
$\tan\beta$, the $t$-quark Yukawa coupling $y_t$  exhibits 
quasi--infrared fixed point (QFP) behaviour
\ref\Pendleton{B.~Pendleton and G.G.~Ross, \plb 98 (1981) 291\semi
C.T.~Hill, \prd 24 (1981) 691}, and the corresponding form  of
$y_t(\mu)$ is favourable for $b/\tau$ unification at $M_G$. For large 
$\tan\beta$, trinification ($y_t(M_G) = y_b(M_G) = y_{\tau} (M_G)$) is
possible\ref\shafi{B.~Ananthanarayan, G.~Lazarides and Q.~Shafi, 
\prd 44 (1991) 1613\semi 
H.~Arason et al, \prl 67 (1991) 2933\semi
V.~Barger, M.S.~Berger and P.~Ohmann, \prd 47 (1993) 1093\semi
M.~Carena et al, \npb 426 (1994) 269\semi
L.J.~Hall, R.~Rattazzi and U.~Sarid, \prd 50 (1994) 7048\semi
R.~Rattazzi and U.~Sarid, \prd 53 (1996) 1553\semi
E.G.~Floratos, G.K.~Leontaris and S.~Lola, \plb 365 (1996) 149}.  
Contemporary analyses generally employ the two-loop gauge and Yukawa 
$\beta$-functions, and apply one loop corrections in the low energy 
theory. In general the change in low energy predictions resulting 
from use of two-loop rather than one-loop $\beta$-functions is quite 
small; it appears that perturbation theory is reliable.
 In this paper we take the first step beyond these     calculations  by 
presenting the three loop $\beta$-functions for the dimensionless
couplings.  We deduce the three-loop Yukawa $\beta$-functions from the
recent  dimensional reduction 
(DRED) calculation of the three loop anomalous dimension of the chiral
supermultiplet in a general $N=1$ theory
\ref\jjn{I.~Jack, D.R.T.~Jones and C.G.~North, 
hep-ph/9603386}. We also need the three-loop gauge $\beta$-functions. 
In Ref.~\ref\jjnb{I.~Jack, D.R.T.~Jones and C.G.~North, hep-ph/9606323}
it is shown by explicit calculation 
that in the abelian case the DRED result for $\beta_g^{(3)}$ 
differs from that obtained from the exact NSVZ formula of 
Ref.\ref\nov{V.A.~Novikov et al, 
\plb166 (1986) 329\semi
M.A.~Shifman and  A.I~Vainstein,
 \npb 277 (1986) 456} 
by a simple coupling constant redefinition, and the corresponding 
redefinition for the non-abelian case is inferred. From Ref.~\jjnb\ we 
are thus able to obtain the DRED results for the three-loop gauge 
$\beta$-functions. In fact the effect of using these rather than 
the results of Ref.~\nov\ is very small in the examples we present here. 

We run gauge and Yukawa couplings between $M_Z$ and $M_G$ 
and compare the results with the corresponding calculations at 
one and two loops. In general the three-loop effects are small and 
opposite in sign to the two-loop ones; certainly 
not more significant than  one- and two-loop radiative corrections 
at $M_Z$ (some of which are not yet calculated). We do not, therefore, 
perform  detailed phenomenological analysis.
The relevant part of the SSM superpotential is: 
\eqn\aaa{W = H_2 t Y_t Q  + H_1 b Y_b Q + H_1 \tau Y_{\tau} L}
where $Y_t$, $Y_b$, $Y_{\tau}$ are $n_g\times n_g$ Yukawa matrices,  
and we define 
\eqn\aaaa{T = Y_t^{\dagger}Y_t^{\phantom {\dagger}}, 
B = Y_b^{\dagger}Y_b^{\phantom {\dagger}}, 
E = Y_{\tau}^{\dagger}Y_{\tau}^{\phantom {\dagger}}, 
\Ttilde = Y_t^{\phantom {\dagger}}Y_t^{\dagger}, 
\Btilde = Y_b^{\phantom {\dagger}}Y_b^{\dagger}, 
\Etilde = Y_{\tau}^{\phantom {\dagger}}Y_{\tau}^{\dagger}.}
The \smgroup\ gauge $\beta$-functions are as follows: 
\eqn\aab{
\beta_{g_i} = (\lf)^{-1}b_i g_i^3 +
(\lf)^{-2}g_i^3\left(\sum_j b_{ij}g_j^2 - a_i\right) 
+ (\lf)^{-3}\beta_{g_i}^{(3)} + \cdots}
where 
\eqn\tlfa{\eqalign{
b_1 &= 2n_g + \frak{3}{5}, \quad b_2 = 2n_g - 5, \quad b_3 = 2n_g - 9\cr
a_1 &= \frak{26}{5}\tr T + \frak{14}{5}\tr B + \frak{18}{5}\tr E,\quad
a_2 = 6\tr T + 6\tr B + 2\tr E,\quad
a_3 = 4\tr T + 4\tr B\cr}}
and 
\eqn\aad{
b_{ij}= \pmatrix{\frak{38}{15}n_g + \frak{9}{25} &
\frak{6}{5}n_g + \frak{9}{5} & \frak{88}{15}n_g\cr
 & & \cr
\frak{2}{5}n_g +\frak{3}{5} & 14n_g -17 & 8n_g\cr
& & \cr
\frak{11}{15}n_g & 3n_g & \frak{68}{3}n_g -54\cr}.
} 

The three loop terms are\jjnb:
\eqna\eva$$\eqalignno{
\beta_{g_1}^{(3)}=&g_1^3\Bigl[\frak{84}{5}\tr T^2+18(\tr T)^2 
+\frak{54}{5}\tr B^2 +\frak{36}{5}(\tr B)^2+\frak{58}{5}\tr TB
+\frak{54}{5}\tr E^2 +\frak{24}{5}(\tr E)^2\cr&+\frak{84}{5}\tr E\tr B
-\left(\frak{169}{75}g_1^2+\frak{87}{5}g_2^2 + \frak{352}{15}g_3^2\right)\tr T
-\left(\frak{49}{75}g_1^2 + \frak{33}{5}g_2^2
+\frak{256}{15}g_3^2\right)\tr B\cr&
-(\frak{81}{25}g_1^2 + \frak{63}{5}g_2^2)\tr E
-(\frak{88}{5}n_g^2 -\frak{572}{9}n_g)g_3^4
-(\frak{18}{5}n_g^2-\frak{9}{5}n_g-\frak{54}{5})g_2^4\cr&
- (\frak{38}{5}n_g^2 +\frak{1261}{225}n_g + \frak{54}{125})g_1^4
-\frak{1096}{225}n_g g_3^2g_1^2-\frak{8}{5}n_g g_2^2g_3^2
-(\frak{27}{25}+ \frak{14}{25}n_g)g_1^2g_2^2\Bigr],
&\eva a\cr
\beta_{g_2}^{(3)}=&g_2^3\Bigl[24\tr T^2+18(\tr T)^2+24\tr B^2
+18(\tr B)^2+12\tr BT+12\tr B\tr E\cr&
+8\tr E^2+2(\tr E)^2-\left(32g_3^2+33g_2^2+\frak{29}{5}g_1^2\right)\tr T
-\left(32g_3^2+33g_2^2+\frak{11}{5}g_1^2\right)\tr B
\cr&-\left(11g_2^2+\frak{21}{5}g_1^2\right)\tr E
-(24n_g^2 -\frak{260}{3}n_g)g_3^4-(26n_g^2-123n_g 
+100)g_2^4
\cr&-(\frak{6}{5}n_g^2 +\frak{169}{75}n_g+\frak{18}{25})g_1^4+
(\frak{2}{5}n_g +\frak{3}{5})g_1^2g_2^2
+8n_g g_2^2g_3^2-\frak{8}{15}n_g g_1^2g_3^2\Bigr],
&\eva b\cr
\beta_{g_3}^{(3)}=&g_3^3\Bigl[18(\tr T)^2+12\tr T^2 +8\tr BT
+12(\tr B)^2+18\tr B^2 +6\tr E\tr B\cr&
-(\frak{104}{3}g_3^2+12g_2^2)(\tr T+\tr B) -g_1^2 (\frak{44}{15}\tr T 
+ \frak{32}{15}\tr B)-(44n_g^2 -\frak{3236}{9}n_g + 567)g_3^4\cr&
+2n_g g_3^2g_2^2+\frak{22}{{45}}n_g g_3^2g_1^2-(\frak{11}{5}n_g^2 
+ \frak{217}{225}n_g)g_1^4
-(9n_g^2-18n_g)g_2^4-\frak{1}{5}n_g g_1^2g_2^2\Bigr].
&\eva c\cr}
$$

For the anomalous dimensions of the chiral superfields we have 
at one loop:

\eqn\tlfb{\eqalign{
\lf\ga^{(1)}_{t} &= 2\Ttilde -\frak{8}{3}g_3^2 -\frak{8}{15}g_1^2,\cr
\lf\ga^{(1)}_{b} &= 2\Btilde -\frak{8}{3}g_3^2 -\frak{2}{15}g_1^2,\cr
\lf\ga^{(1)}_{Q} &= B + T - \frak{8}{3}g_3^2 - \frak{3}{2}g_2^2
-\frak{1}{30}g_1^2,\cr
\lf\ga^{(1)}_{H_2} &= 3\tr T - \frak{3}{2}g_2^2 -\frak{3}{10}g_1^2,\cr
\lf\ga^{(1)}_{H_1} &= \tr E +3\tr B-\frak{3}{2}g_2^2 
-\frak{3}{10}g_1^2,\cr
\lf\ga^{(1)}_{L} &= E - \frak{3}{2}g_2^2
-\frak{3}{10}g_1^2,\cr
\lf\ga^{(1)}_{\tau} &= 2\Etilde-\frak{6}{5}g_1^2.\cr}}
and at two loops\ref\bjork{J.E.~Bj\"orkman and D.R.T.~Jones, 
\npb 259 (1985) 533}:

\eqna\Keva$$\eqalignno{
\llf\gamma_t^{(2)}&=-2\Ttilde^2-6(\tr T)\Ttilde
-2Y_t^{\phantom {\dagger}} BY_t^{\dagger}
+\left(6g_2^2-\frak{2}{5}g_1^2\right)\Ttilde
\cr&+ (\frak{8}{15}b_1 +\frak{64}{{225}})g_1^4
+\frak{128}{{45}}g_1^2g_3^2 + (\frak{8}{3}b_3 + \frak{64}{9})g_3^4,
&\Keva a\cr
\llf\gamma_b^{(2)}&=-2\Btilde^2-6(\tr B)\Btilde
-2Y_b^{\phantom {\dagger}} TY_b^{\dagger}-2(\tr E)\Btilde
+\left(6g_2^2+\frak{2}{5}g_1^2\right)\Btilde\cr&
+(\frak{2}{15}b_1 +\frak{4}{{225}})g_1^4
+\frak{32}{{45}}g_1^2g_3^2 +(\frak{8}{3}b_3 + \frak{64}{9})g_3^4,
&\Keva b\cr
\llf\gamma_Q^{(2)}&=-2T^2-3(\tr T)T-2B^2
-3(\tr B)B
-(\tr E)B +g_1^2(\frak{4}{5}T+\frak{2}{5}B)
+\frak{1}{{10}}g_1^2g_2^2\cr&
+8g_3^2g_2^2+\frak{8}{{45}}g_1^2g_3^2
+(\frak{8}{3}b_3 + \frak{64}{9})g_3^4
+(\frak{3}{2}b_2 + \frak{9}{4})g_2^4
+ (\frak{1}{30}b_1 + \frak{1}{{900}})g_1^4,
&\Keva c\cr
\llf\gamma_{H_2}^{(2)}&=-9\tr T^2 -3\tr BT
+\left(16g_3^2+\frak{4}{5}g_1^2\right)\tr T
+(\frak{3}{2}b_2 + \frak{9}{4})g_2^4+\frak{9}{{10}}g_1^2g_2^2\cr&
+(\frak{3}{10}b_1 + \frak{9}{100})g_1^4,
&\Keva d\cr
\llf\gamma_{H_1}^{(2)}&=-9\tr B^2 -3\tr BT-3(\tr E^2)
+\left(16g_3^2-\frak{2}{5}g_1^2\right)\tr B
+\frak{6}{5}g_1^2\tr E\cr
&+(\frak{3}{2}b_2 + \frak{9}{4})g_2^4+\frak{9}{{10}}g_1^2g_2^2
+(\frak{3}{10}b_1 + \frak{9}{100})g_1^4,
&\Keva e\cr
\llf\gamma_L^{(2)}&=-2E^2 -3(\tr B)E-(\tr E)E
+\frak{6}{5}g_1^2E+(\frak{3}{2}b_2 + \frak{9}{4})g_2^4
\cr&+\frak{9}{{10}}g_1^2g_2^2+(\frak{3}{10}b_1 + \frak{9}{100})g_1^4,
&\Keva f\cr
\llf\gamma_{\tau}^{(2)}&=-2\Etilde^2-6(\tr B)\Etilde-2(\tr E)\Etilde
+\left(6g_2^2-\frak{6}{5}g_1^2\right)\Etilde+(\frak{6}{5}b_1 
+ \frak{36}{{25}})g_1^4.
&\Keva g\cr}$$

The three loop results are\jjn: 
\eqna\Kevb$$\eqalignno{
\lllf\gamma^{(3)}_t &= (2\kappa +6)\Ttilde^3+36(\tr T^2)\Ttilde
+6(\tr T)\Ttilde^2-18(\tr T)^2\Ttilde
+12(\tr TB)\Ttilde\cr&
-2Y_t^{\phantom {\dagger}} BTY_t^{\dagger}
-2Y_t^{\phantom {\dagger}} TBY_t^{\dagger}
-6(\tr T)Y_t^{\phantom {\dagger}} BY_t^{\dagger}
+12(\tr B)Y_t^{\phantom {\dagger}} BY_t^{\dagger}
+6Y_t^{\phantom {\dagger}} B^2Y_t^{\dagger}\cr&
+4(\tr E)Y_t^{\phantom {\dagger}} BY_t^{\dagger}
+\left[\frak{64}{3}g_3^2+(9-3\kappa )g_2^2
+\left(\kappa -\frak{1}{3}\right)g_1^2\right]\Ttilde^2
+\Bigl[\left(16\kappa -16\right)g_3^2\cr&-(9\kappa -27)g_2^2
+\left(7+\frak{7}{5}\kappa \right)g_1^2\Bigr](\tr T)\Ttilde
+\left[\frak{64}{3}g_3^2-(3\kappa -9)g_2^2+\left(\frak{19}{{15}}
+\frak{3}{5}\kappa \right)g_1^2\right]Y_t^{\phantom {\dagger}}BY_t^{\dagger}\cr
&+\Bigl[\left(\frak{304}{3}-\frak{272}{9}\kappa -32n_g\right)g_3^4 
+(16\kappa -88)g_2^2g_3^2-\left(\frak{112}{{45}}\kappa
+\frak{8}{{15}}\right)g_1^2g_3^2
\cr&-\left(\frak{3}{2}\kappa +24n_g-\frak{57}{2}\right)g_2^4
-\left(\frak{67}{5}-\frak{13}{5}\kappa \right)g_1^2g_2^2
-\left(\frak{247}{{450}}\kappa +\frak{237}{{150}}
+\frak{24}{5}n_g\right)g_1^4\Bigr]\Ttilde
\cr&-\left(\frak{80}{3}g_3^4+\frak{104}{{15}}g_1^4\right)\tr T
-\left(\frak{56}{{15}}g_1^4+\frak{80}{3}g_3^4\right)\tr B
-\frak{24}{5}g_1^4\tr E
\cr&+\left(\frak{160}{9}\kappa n_g+\frak{32}{3}n_g^2+\frak{368}{9}n_g
-\frak{3184}{{27}}\right)g_3^6
-\left(4\kappa n_g-20n_g\right)g_2^2g_3^4\cr&
+\left(\frak{92}{{45}}n_g-\frak{44}{{45}}\kappa n_g
-\frak{448}{{45}}\right)g_1^2g_3^4
-\left(\frak{352}{{225}}\kappa n_g
+\frak{1216}{{225}}-\frak{224}{{45}}n_g\right)g_1^4g_3^2
-\Bigl[\left(\frak{12}{{25}}+\frak{8}{{25}}n_g\right)\kappa\cr& 
-\frak{8}{5}n_g-\frak{12}{5}\Bigr]g_1^4g_2^2
-\Bigl[\left(\frak{12}{{125}}+\frak{152}{{225}}n_g\right)\kappa 
-\frak{184}{{45}}n_g-\frak{32}{{15}}n_g^2-\frak{668}{{3375}}\Bigr]g_1^6,
&\Kevb a\cr
\lllf\gamma^{(3)}_b &= (2\kappa +6)\Btilde^3 + 36(\tr B^2)\Btilde
+6(\tr B)\Btilde^2 - 18(\tr B)^2\Btilde+12(\tr BT)\Btilde\cr&
-2Y_b^{\phantom {\dagger}} TBY_b^{\dagger}
-2Y_b^{\phantom {\dagger}} BTY_b^{\dagger}
-6(\tr B)Y_b^{\phantom {\dagger}} TY_b^{\dagger}
+12(\tr T)Y_b^{\phantom {\dagger}} TY_b^{\dagger}
+6Y_b^{\phantom {\dagger}} T^2Y_b^{\dagger}\cr&+12(\tr E^2)\Btilde
+2(\tr E)\Btilde^2 - 2(\tr E)Y_b^{\phantom {\dagger}} TY_b^{\dagger}
-12(\tr E)(\tr B)\Btilde
-2(\tr E)^2\Btilde\cr&+\Bigl[\frak{64}{3}g_3^2-(3\kappa -9)g_2^2+(
\frak{3}{5}\kappa 
-\frak{29}{{15}})g_1^2\Bigr]Y_b^{\phantom {\dagger}} TY_b^{\dagger}
+\Bigl[\frak{64}{3}g_3^2-(3\kappa -9)g_2^2\cr&
+\left(\frak{1}{5}\kappa-\frak{1}{3}\right)g_1^2   
\Bigr]Y_b^{\phantom {\dagger}} BY_b^{\dagger}
+\Bigl[\left(16\kappa -16\right)g_3^2  -(9\kappa -27)g_2^2
+\left(7 -\kappa \right)g_1^2\Bigr](\tr B)\Btilde\cr&
+\Bigl[16g_3^2+(9 -3\kappa )g_2^2 +(\kappa -3)g_1^2\Bigr](\tr E)\Btilde
+\Bigl[-\left(\frak{272}{9}\kappa -\frak{304}{3}+32n_g\right)g_3^4\cr&
+\left(16\kappa -88\right)g_2^2g_3^2+\left(\frak{16}{9}\kappa
-\frak{24}{5}\right)g_1^2g_3^2
-\left(\frak{3}{2}\kappa +24n_g-\frak{57}{2}\right)g_2^4+
\left(\frak{7}{5}\kappa -\frak{43}{5}\right)g_1^2g_2^2\cr&
-\left(\frak{49}{{30}}+\frak{16}{5}n_g
+\frak{7}{{450}}\kappa \right)g_1^4\Bigr]\Btilde
-\left(\frak{26}{15}g_1^4+\frak{80}{3}g_3^4\right)\tr T
-\Bigl[\frak{80}{3}g_3^4+\frak{14}{{15}}g_1^4\Bigr]\tr B\cr&
-\frak{6}{5}g_1^4\tr \Etilde
+\left(\frak{160}{9}\kappa n_g+\frak{32}{3}n_g^2+\frak{368}{9}n_g
-\frak{3184}{{27}}\right)g_3^6
-\Bigl[\left(\frak{3}{{25}}
+\frak{2}{{25}}n_g\right)\kappa -\frak{2}{5}n_g\cr&
-\frak{3}{5}\Bigr]g_1^4g_2^2
-\left(\frak{112}{{45}}+\frak{44}{{45}}\kappa n_g
-\frak{188}{{45}}n_g\right)g_1^2g_3^4
-\left(\frak{88}{{225}}\kappa n_g-\frak{56}{{45}}n_g
+\frak{112}{{225}}\right)g_1^4g_3^2\cr&
-(4\kappa n_g-20n_g)g_2^2g_3^4-\Bigl[\left(\frak{38}{{225}}n_g
+\frak{3}{{125}}\right)\kappa -\frak{8}{{15}}n_g^2-\frak{254}{{225}}n_g
-\frak{103}{{675}}\Bigr]g_1^6,
&\Kevb b\cr
\lllf\gamma^{(3)}_Q &= \kappa T^3+18(\tr T^2)T+6(\tr T)T^2
-9(\tr T)^2T+\kappa B^3+18(\tr B^2)B
+6(\tr B)B^2\cr&-9(\tr B)^2B+6(\tr BT)T 
+4TBT+6(\tr TB)B+4BTB
+6(\tr E^2)B-(\tr E)^2B\cr&
 +2(\tr E)B^2  -6(\tr B)(\tr E)B
+\Bigl[\frak{64}{3}g_3^2+(3\kappa -3)g_2^2
-\left(\kappa -\frak{11}{3}\right)g_1^2\Bigr]T^2\cr&
+\Bigl[(8\kappa -8)g_3^2+18g_2^2-\left(\frak{4}{5}\kappa 
-2\right)g_1^2\Bigr](\tr T)T
+\Bigl[\frak{64}{3}g_3^2+(3\kappa -3)g_2^2
-\left(\frak{1}{5}\kappa-\frak{7}{15}\right)g_1^2\Bigr]B^2\cr&
+\Bigl[(8\kappa -8)g_3^2+18g_2^2-\left(\frak{4}{5}\kappa 
-\frak{16}{5}\right)g_1^2\Bigr](\tr B)B
+\Bigl[8g_3^2+6g_2^2+\left(\frak{2}{5}\kappa -\frak{8}{5}
 \right)g_1^2\Bigr](\tr E)B\cr&
+\Bigl[-\left(\frak{136}{9}\kappa +16n_g-\frak{152}{3}\right)g_3^4
-4g_2^2g_3^2-\left(\frak{68}{5}-\frak{64}{{45}}\kappa \right)g_1^2g_3^2
-\left(9n_g+\frak{21}{4}\kappa -\frak{63}{4}\right)g_2^4\cr&
+\left(\frak{3}{2}\kappa-\frak{59}{{10}}
 \right)g_1^2g_2^2
+\left(\frak{143}{{900}}\kappa -\frak{707}{{300}}
-\frak{17}{5}n_g\right)g_1^4\Bigr]T
+\Bigl[-\left(\frak{136}{9}\kappa +16n_g-\frak{152}{3}\right)g_3^4\cr&
-4g_2^2g_3^2+\left(\frak{64}{{45}}\kappa -\frak{76}{{15}}\right)g_1^2g_3^2
-\left(9n_g+\frak{21}{4}\kappa -\frak{63}{4}\right)g_2^4
+\left(\frak{3}{{10}}\kappa -\frak{41}{{10}}\right)g_1^2g_2^2\cr&
-\left(\frak{9}{5}n_g+\frak{279}{300}-\frak{7}{180}\kappa \right)g_1^4\Bigr]B
-\Bigl[\frak{80}{3}g_3^4+\frak{45}{2}g_2^4
+\frak{13}{{30}}g_1^4\Bigr]\tr T\cr&
-\Bigl[\frak{80}{3}g_3^4+\frak{45}{2}g_2^4
+\frak{7}{{30}}g_1^4\Bigr]\tr B
-\left(\frak{15}{2}g_2^4+\frak{3}{10}g_1^4\right)\tr E
+\Bigl[\frak{160}{9}n_g\kappa +\frak{32}{3}n_g^2\cr&
+\frak{368}{9}n_g-\frak{3184}{27}\Bigr]g_3^6
+\left(12n_g-4\kappa n_g-28\right)g_2^2g_3^4
+\left(\frak{212}{45}n_g-\frak{44}{45}\kappa n_g
-\frak{28}{45}\right)g_1^2g_3^4\cr&   
-\Bigl[\left(\frak{3}{100}+\frak{1}{50}n_g\right)\kappa 
-\frak{11}{100}\Bigr]g_1^4g_2^2-\frak{8}{5}g_1^2g_2^2g_3^2
+\left(\frak{14}{45}n_g-\frak{22}{225}\kappa n_g
-\frak{16}{225}\right)g_1^4g_3^2
\cr&+\Bigl[\left(\frak{15}{2}n_g+\frak{15}{4}\right)\kappa -\frak{87}{4}
+18n_g+6n_g^2\Bigr]g_2^6
-\Bigl[\left(\frak{9}{20}+\frak{3}{10}n_g\right)\kappa -\frak{7}{5}n_g
-\frak{41}{20}\Bigr]g_1^2g_2^4
\cr&-\left(6\kappa n_g-22n_g+16\right)g_2^4g_3^2
-\Bigl[\left(\frak{3}{500}+\frak{19}{450}n_g\right)\kappa -\frak{2}{15}n_g^2
-\frak{13}{45}n_g-\frak{557}{13500}\Bigr]g_1^6,
&\Kevb c\cr
\lllf\gamma_{H_2}^{(3)} &= (3\kappa +3)\tr T^3 +54\tr T\tr T^2 
+9\tr TB^2
+18\tr B\tr BT+6\tr E\tr BT\cr&
+\Bigl[(72-24\kappa )g_3^2+(9+9\kappa )g_2^2+\left(\frak{57}{5}
-\frak{3}{5}\kappa \right)g_1^2\Bigr]
\tr T^2 
+\Bigl[(24-8\kappa)g_3^2+18g_2^2\cr&+\left(\frak{1}{5}\kappa 
+\frak{6}{5}\right)g_1^2\Bigr]\tr BT
-\Bigl[\left(64n_g+\frak{8}{3}\kappa -\frak{416}{3}\right)g_3^4
+\left(27n_g - \frak{99}{4}  
+\frak{63}{4}\kappa \right)g_2^4\cr&
+\left(\frak{43}{5}n_g+\frak{115}{12}
+\frak{13}{60}\kappa \right)g_1^4
-\left(24\kappa -132\right)g_2^2g_3^2
-\left(\frak{104}{15}\kappa -\frak{124}{3}\right)g_1^2g_3^2 
\cr&-\left(\frak{21}{10}\kappa -\frak{57}{10}\right)g_1^2g_2^2
\Bigr]\tr T
-\left(\frak{45}{2}g_2^4+\frak{21}{10}g_1^4\right)\tr B
-\left(\frak{15}{2}g_2^4+\frak{27}{10}g_1^4\right)\tr E+\Xi, 
&\Kevb d\cr
\lllf\gamma_{H_1}^{(3)} &= (3\kappa +3)\tr B^3 +54\tr B\tr B^2 
+18\tr T\tr BT+9\tr BT^2+(\kappa +1)\tr E^3
\cr&
+6\tr E\tr E^2+18\tr B\tr E^2+18\tr E\tr B^2 
+\Bigl[\left(\frak{7}{5}\kappa -\frak{12}{5}\right)g_1^2
\cr&
+(24-8\kappa )g_3^2+18g_2^2\Bigr]\tr BT
+\Bigl[(72-24\kappa )g_3^2+(9+9\kappa )g_2^2
+\left(\frak{9}{5}\kappa +3\right)g_1^2\Bigr]\tr B^2 
\cr&
+\Bigl[\left(3\kappa +3\right)g_2^2-\left(\frak{9}{5}\kappa 
-9\right)g_1^2\Bigr](\tr E^2)
-\left(\frak{45}{2}g_2^4+\frak{39}{10}g_1^4\right)\tr T
+\Bigl[\left(\frak{416}{3}-64n_g-\frak{8}{3}\kappa \right)g_3^4\cr&
+\left(24\kappa -132\right)g_2^2g_3^2
+\left(\frak{56}{15}\kappa-\frak{284}{15}\right)g_1^2g_3^2
-\left(\frak{63}{4}\kappa -\frak{99}{4}+27n_g\right)g_2^4\cr&
-\left(\frak{3}{2}\kappa +\frak{3}{10}\right)g_1^2g_2^2
-\left(\frak{77}{300}\kappa +\frak{191}{60}
+\frak{19}{5}n_g\right)g_1^4\Bigr]\tr B
+\Bigl[\left(\frak{33}{4}-\frak{21}{4}\kappa -9n_g\right)g_2^4\cr&
+\left(\frak{27}{10}\kappa-\frak{81}{10}\right)g_1^2g_2^2
+\left(\frak{27}{100}\kappa -\frak{33}{5}n_g
-\frak{207}{20}\right)g_1^4\Bigr]\tr E+\Xi,
&\Kevb e\cr
\lllf\gamma_L^{(3)} &= \kappa E^3+
6(\tr E^2)E+2(\tr E)E^2
-(\tr E)^2E+6(\tr B)E^2-6(\tr B)(\tr E)E
\cr&+18(\tr B^2)E+6(\tr BT)E-9(\tr B)^2E
+\Bigl[(3\kappa -3)g_2^2-\left(\frak{9}{5}\kappa -9\right)g_1^2\Bigr]E^2
\cr&+\Bigl[(8\kappa -32)g_3^2+18g_2^2
-(2\kappa -8)g_1^2\Bigr](\tr B)E
+6g_2^2(\tr E)E
-\left(\frak{39}{10}g_1^4+\frak{45}{2}g_2^4\right)\tr T
\cr&+\Bigl[\left(\frak{63}{4}-\frak{21}{4}\kappa -9n_g\right)g_2^4
+\left(\frak{27}{10}\kappa -\frak{81}{10}\right)g_1^2g_2^2
+\left(\frak{27}{100}\kappa -\frak{153}{20} 
-\frak{33}{5}n_g\right)g_1^4\Bigr]E
\cr&-\Bigl[\frak{21}{{10}}g_1^4+\frak{45}{2}g_2^4\Bigr]\tr B
-\Bigl[\frak{15}{2}g_2^4+\frak{27}{{10}}g_1^4\Bigr]\tr E
+\Xi, 
&\Kevb f\cr
\lllf\gamma_{\tau}^{(3)} &
= (2\kappa +6)\Etilde^3+12(\tr E^2)\Etilde
+2(\tr E)\Etilde^2-2(\tr E)^2\Etilde+6(\tr B)\Etilde^2\cr&
-12(\tr B)(\tr E)\Etilde+36(\tr B^2)\Etilde-18(\tr B)^2\Etilde
+12(\tr BT)\Etilde+\Bigl[(9-3\kappa)g_2^2\cr&
+(\frak{9}{5}\kappa +\frak{9}{5})g_1^2\Bigr]\Bigl[\Etilde^2
+(\tr E)\Etilde\Bigr]
+\Bigl[\left(16\kappa -64\right)g_3^2
-\left(9\kappa -27\right)g_2^2
\cr&
+\left(\frak{107}{5}
+\frak{7}{5}\kappa \right)g_1^2\Bigr](\tr B)\Etilde
-\frak{78}{5}g_1^4\tr T
-\frak{42}{5}g_1^4\tr B-\frak{54}{5}g_1^4\tr E
\cr&+\Bigl[-\left(\frak{3}{2}\kappa +24n_g-\frak{57}{2}\right)g_2^4
+\left(\frak{27}{5}\kappa -27\right)g_1^2g_2^2
-\left(\frak{63}{50}+\frak{27}{10}\kappa
+\frak{48}{5}n_g\right)g_1^4\Bigr]\Etilde
\cr&+\left(\frak{88}{5}n_g-\frak{88}{25}\kappa n_g\right)g_1^4g_3^2
+\Bigl[\frak{18}{5}n_g+\frak{27}{5}
-\left(\frak{27}{25}+\frak{18}{25}n_g\right)\kappa\Bigr]g_1^4g_2^2
\cr&+\Bigl[\frak{24}{5}n_g^2+\frak{38}{5}n_g
-\left(\frak{38}{25}n_g+\frak{27}{125}\right)\kappa
-\frak{351}{125}\Bigr]g_1^6, 
&\Kevb g\cr}$$
where $\kappa  = 6\zeta(3)$, and 
\eqn\tlfd{\eqalign{
\Xi &= n_g\left(30-6\kappa \right)g_2^4g_3^2+n_g\left(\frak{22}{5}
-\frak{22}{25}\kappa \right)g_1^4g_3^2
+\Bigl[\left(\frak{15}{2}n_g+\frak{15}{4}\right)\kappa -\frak{87}{4}
+18n_g+6n_g^2\Bigr]g_2^6\cr&
-\Bigl[\left(\frak{9}{20}+\frak{3}{10}n_g\right)\kappa 
-\frak{9}{20}-\frak{3}{5}n_g\Bigr]g_1^2g_2^4
-\Bigl[\left(\frak{9}{50}n_g+\frak{27}{100}\right)\kappa 
-\frak{27}{100}\Bigr]g_1^4g_2^2\cr&
-\Bigl[\left(\frak{19}{50}n_g+\frak{27}{500}\right)\kappa 
-\frak{27}{100}-\frak{6}{5}n_g^2
-\frak{61}{25}n_g\Bigr]g_1^6.\cr}}

In terms of the anomalous dimensions, the Yukawa $\beta$-functions are:
\eqn\tlfb{
\beta_{Y_t} = \gamma_t Y_t + Y_t ( \gamma_Q + \gamma_{H_2})
,\quad
\beta_{Y_b} = \gamma_b Y_b + Y_b ( \gamma_Q + \gamma_{H_1})
,\quad
\beta_{Y_{\tau}} = \gamma_{\tau} Y_{\tau} 
+ Y_{\tau} ( \gamma_L + \gamma_{H_1}).
}
In the approximation that we retain only $\alpha_3$ and the $t$-quark 
Yukawa coupling $y_t$ then we have (for $n_g = 3$): 
\eqna\Kevd$$\eqalignno{
\lf\beta_{y_t}^{(1)} &= y_t (6y_t^2 - \frak{16}{3}g_3^2), 
&\Kevd a\cr
\llf\beta_{y_t}^{(2)} &= y_t (-22y_t^4 + 16y_t^2g_3^2 - \frak{16}{9}g_3^4),
&\Kevd b\cr
\lllf\beta_{y_t}^{(3)} &= y_t ([102+ 6\kappa ]y_t^6 
+ \frak{272}{3}y_t^4g_3^2  - [\frak{296}{3}+48\kappa ]y_t^2g_3^4
  +[\frak{5440}{27} + \frak{320}{3}\kappa ]g_3^6). 
&\Kevd c\cr}$$
We see that for values of $y_t$ in the neighbourhood of or greater than 
the one-loop QFP 
(which corresponds to $y_t\approx 1$) we have 
$\beta_{y_t}^{(2)}< 0$ and $\beta_{y_t}^{(3)}>0$. We may therefore expect 
that where three loop contributions are not completely negligible they     
will tend to cancel the two loop contributions. We will see examples of 
this behaviour presently.

\epsfysize= 2.5in
\centerline{\epsfbox{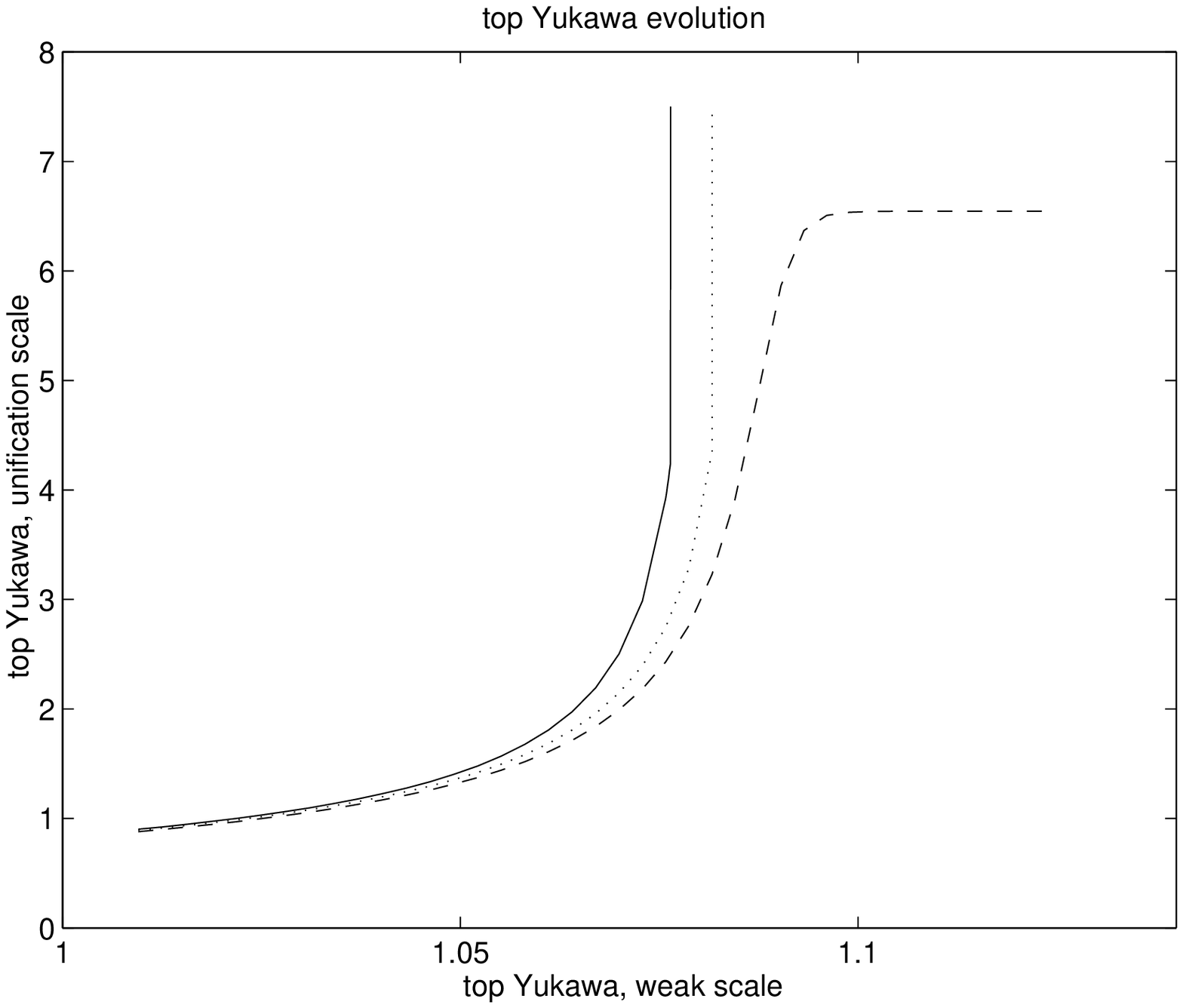}}
\in
{\it \noindent Fig.1:
Plot of $y_t(M_G)$ against $y_t(M_Z)$.
The solid, dashed and dotted lines correspond to one, two and 
three-loop $\beta$-functions respectively.}
\medskip
\out
For small $\tan\beta$, QFP behaviour is of interest  because there is a
large domain of possible values of $y_t(M_G)$ that  lead to the same
value  of $y_t (M_Z)$. (For a recent discussion, see 
Ref.~\ref\lanz{M.~Lanzagorta and G.G.~Ross, \plb 349 (1995) 319}.) 
It is important, however,   to consider to what
extent this domain is restricted by the requirement of  perturbative
believability. Setting $g_3 = 0$ above, we see that
$|\beta_{Y_t}^{(2)}/\beta_{Y_t}^{(1)}|\approx 1$ for $y_t\approx 6.6$
while $|\beta_{Y_t}^{(3)}/\beta_{Y_t}^{(2)}|\approx 1$ for $y_t\approx
4.9$.  So since the QFP   corresponds to $y_t\approx 1.1$, we may expect
there to be  a good sized domain available for $y_t (M_G)$ such that we
have both  perturbative believability and approach to the QFP at low
energies.  This is illustrated in Figure 1, where 
(with $m_t^{\rm pole} = 175$GeV, and using the complete $\beta$-functions, 
not the Eq.~\Kevd{}\ approximations) we plot $y_t(M_G)$
against  $y_t(M_Z)$, for values close to the QFP. The breakdown of
perturbation  theory at $y_t (M_G)\approx 6$ is clearly seen.  

In our running analysis
\footnote{\dag}{For a different approach 
to this running analysis based on the NSVZ $\beta_g$, 
see Ref.~\ref\shif{M.~Shifman, hep-ph/9606281}.}  
we take the effective field theory  to be the
SSM for all scales between $M_Z$ and $M_G$: see 
Ref.~\ref\chank{A.E.~Faraggi and B.~Grinstein, \npb 422 (1994) 3\semi
P.H.~Chankowski, Z.~Pluciennik and S.~Pokorski, \npb 439 (1995) 23\semi
J.~Bagger, K.~Matchev and  D.~Pierce, \plb 348 (1995) 443\semi
D.~Pierce et al, hep-ph/9606211}\  for a
discussion of this procedure. A reason to prefer this to the
traditional stepwise method is that the latter  involves non-\sic\
intermediate theories for which (beyond one loop) use  of DRED rather
than DREG is problematic \ref\jjr{I.~Jack, D.R.T.~Jones and
K.L.~Roberts,  \zpc 62 (1994) 161;  {\it ibid}\/{\bf C}63  (1994) 151}.
This choice, means, of course,  that the input values of $\alpha_{1\to
3}$ are sensitive to our  assumptions about the sparticle spectrum. We
have corrected for this at one-loop  using Ref.~\chank;  of course we
should use two-loop  corrections for consistency, as we should, for
example,  in converting the running $m_t$ to the pole $m_t$. These
calculations remain to be done however;  there exists a result for the 
two-loop gluon contribution to $m_t^{\rm pole}$ but  this is in DREG not
DRED. We do not impose trinification  of the gauge couplings, since 
this leads to somewhat high values  of $\alpha_3 (M_Z)$; for a
discussion and references see  Ref.~\ref\roszk{L.~Roszkowski,
hep-ph/9509273}.  We input $\alpha_3$ and define unification to 
be where $\alpha_1$ and $\alpha_2$ meet. Our input values at $M_Z$ are
$\alpha_1 = 0.0167$, $\alpha_2 = 0.032$ and  $\alpha_3 = 0.1$. These
values correspond to    a  superpartner spectrum with an effective
supersymmetric  scale\ref\carena{ P.~Langacker and N.~Polonski, \prd 47
(1993) 4028\semi M.~Carena, S.~Pokorski and C.E.M.~Wagner, \npb 406
(1993) 59}of   $T_{\rm SUSY} = 1$TeV, and $\alpha_3^{\rm SM} = 0.117$.

\epsfysize= 2.5in
\centerline{\epsfbox{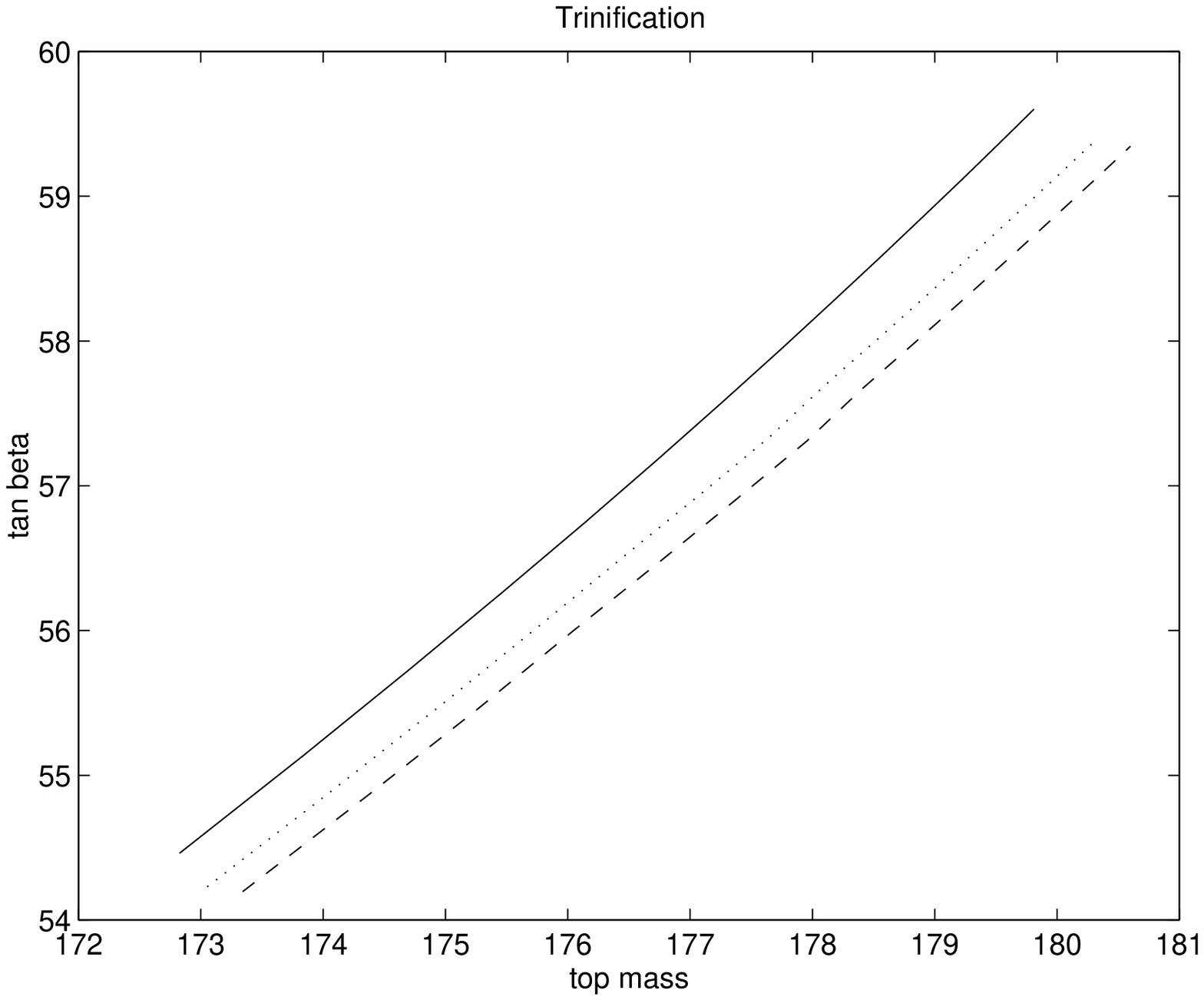}}
\in
{\it \noindent Fig.2:
Plot of $\tan\beta$ against $m_t^{\rm pole}$ with Yukawa trinification.
The solid, dashed and dotted lines correspond to one, two and 
three-loop $\beta$-functions respectively.}
\medskip
\out

In Figure~2 we plot $\tan\beta$ against $m_t$ assuming trinification
($y_t = y_b = y_{\tau}$) at $M_G$. As anticipated, the three loop
corrections counteract (to some extent)  the two loop ones. Their effect, 
while small, is not completely negligible. Proximity of $y_t$ and $y_b$ to 
their QFPs leads to an upper limit $m_t\approx 181$GeV.    
(In comparing with, for example Fig.~1 of
Ref.~\ref\Carenab{M.~Carena et al,  Ref.~\shafi}, it is important to
note that there they have taken the sparticle  masses to be at $M_Z$.) 

\epsfysize= 2.5in
\centerline{\epsfbox{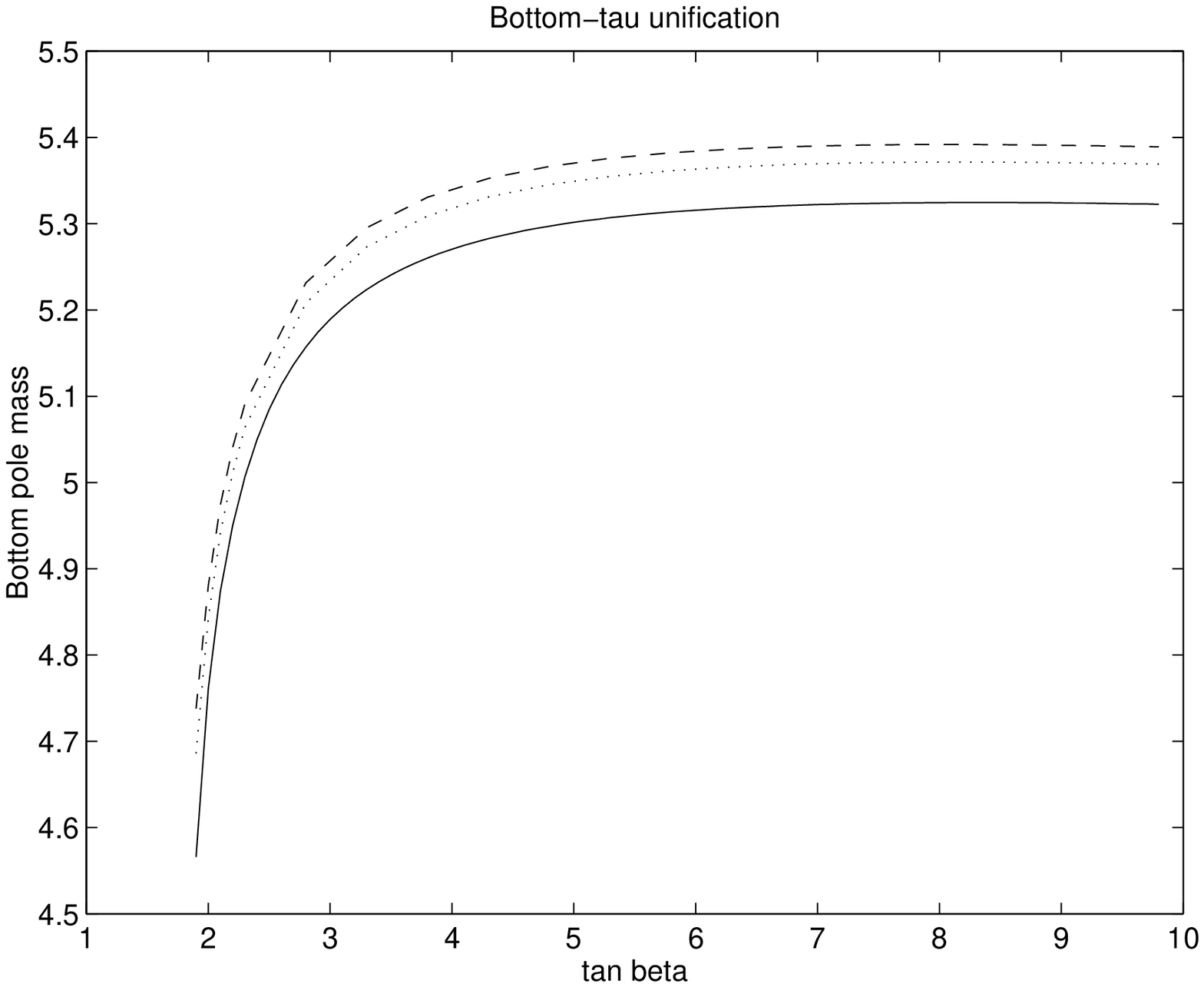}}
\in
{\it \noindent Fig.3:
Plot of $m_b^{\rm pole}$ against $\tan\beta$.
The solid, dashed and dotted lines correspond to one, two and 
three-loop $\beta$-functions respectively.}
\medskip
\out

As another example, consider the low $\tan\beta$ region. In Figure~3  we
plot $m_b^{\rm pole}$ against $\tan\beta$, where we have imposed
$b-\tau$  unification. (For running from $M_Z$ to $m_b$ we use the 
$SU_3\otimes\ U_1$ $\beta$-functions given, for example in 
Ref.~\ref\arasonb{H.~Arason et al, \prd 46 (1992) 3945}.)  In this graph
the QFP is approached  as $\tan\beta$ decreases. 
(The fact that proximity to the QFP gives a better value of 
$m_b^{\rm pole}$ has been noted by a number of authors; 
for an early example see  Ref.~\bjork.) 
The result for $m_b^{\rm pole}$ is quite sensitive to the 
input value of $\alpha_3$ (and hence to the sparticle spectrum).

In conclusion: detailed running coupling analyses for the dimensionless 
SSM couplings have been performed by a number of groups; we have not 
here duplicated in full these efforts,  but instead investigated the
effect on them  of the three loop $\beta$-functions. We have seen that
the corrections are small; nevertheless it is possible that one
day they will play a part  in a   very accurate comparison between the
SSM and  experiment. 

In non-minimal models, the three-loop terms may assume 
more immediate importance. Specifically, consider the possibility 
that $\beta_{g_3}^{(1)}=0$. In such cases two-loop corrections 
will clearly dominate the evolution of $g_3$. It was shown 
in Ref.~\bjork\ that at two loops perturbative unification is only just 
achievable. Evidently three-loop contributions will be important 
here, and in fact improve matters. Models of this kind will 
have interesting phenomenology, with significant differences 
from the 
SSM.\ref\kmr{C.~Kolda and J.~March-Russell, to be published}

\bigskip\centerline{{\bf Acknowledgements}}\nobreak

IJ was supported by PPARC via an Advanced Fellowship, and PF by a 
scholarship from JNICT. We thank Chris Kolda, John March-Russell and 
Damien Pierce for conversations. 

\listrefs
\bye